\documentclass[journal]{IEEEtran}

\usepackage[utf8]{inputenc}
\usepackage[T1]{fontenc}
\usepackage{amsmath,amssymb}
\usepackage[nointegrals]{wasysym}
\usepackage{graphicx}
\usepackage{booktabs,array}
\usepackage{url}
\usepackage{xcolor}
\usepackage{hyperref}
\usepackage{cite}
\usepackage{etoolbox}
\AtBeginDocument{\let\citeorig\cite}
\newcommand{\tablecite}[1]{\mbox{\citeorig{#1}}}
\AtBeginEnvironment{tabular}{\let\cite\tablecite}

\graphicspath{{figures/}}

\begin{document}

\title{Infrastructure for Verifiable AI: Where Proof Costs Go and How Hardware Can Help}
\author{Yihan~Xia, Wenxin~Xu, and Taotao~Wang%
\thanks{\raggedright The authors are with the College of Electronics and Information Engineering, Shenzhen University, Shenzhen, China. E-mail: \texttt{xiayihan2023@email.szu.edu.cn} (Y. Xia); \texttt{2022280280@email.szu.edu.cn} (W. Xu); \texttt{ttwang@szu.edu.cn} (T. Wang). \textit{Corresponding author: T. Wang.}}}
\markboth{Infrastructure for Verifiable AI}{Infrastructure for Verifiable AI}
\maketitle

\begin{abstract}
Users of artificial intelligence services, including those powered by large language models and other transformer-based models, need assurance that inference and training follow the declared computation without exposing private models or data. Zero-knowledge proofs can provide this assurance, but proof generation remains expensive. Hardware acceleration research often reports gains for individual cryptographic operations, leaving their effect on the full cost of proving neural-network workloads, particularly large transformer models, unclear. We introduce a four-layer framework linking verifiable AI workloads, proof systems, cryptographic kernels and hardware platforms, use it to review the literature, and add original measurements of six provers and seven kernels to identify which costs matter in practice. Our measurements show that the dominant cost changes with the proof system and input length. In low-rank fine-tuning, the cost of opening commitments to fixed model weights remains nearly constant across input lengths from 64 to 2048, while the work associated with intermediate values grows. A 13-billion-parameter fine-tuning configuration exceeds the capacity of a 40-gigabyte graphics processor but completes on an 80-gigabyte device, showing why capacity must be assessed alongside execution time. Our kernel profiling further identifies instruction issue and synchronization limits that a bandwidth-only model misses. We use these findings to assess graphics processors, reconfigurable logic, custom chips and existing artificial intelligence accelerators. The resulting research agenda connects proof-friendly numeric formats, reuse of fixed-weight commitments, heterogeneous execution and shared benchmarks to the measured bottlenecks. Effective acceleration requires selecting hardware and proof-system optimizations around the cost profile of the target workload.
\end{abstract}


\section{Introduction}
\label{sec:intro}

\IEEEPARstart{A}{rtificial} intelligence, including large language models and other transformer-based models, is increasingly consumed as a service. Users send prompts to an inference interface, upload private data to a fine-tuning service, or rely on a provider's claim that training followed a declared procedure. In each case the computation happens on someone else's hardware, and the user sees only the result. This leaves room for quiet deviations that are hard to detect from the outside: an inference provider may serve a smaller or more aggressively quantized model than the one it advertises~\cite{svip,toploc}; a fine-tuning service may skip part of the user's data or use different hyper-parameters~\cite{aftune,vft}; a training pipeline may omit declared preprocessing, optimization, or privacy-preserving steps~\cite{veridp,pol}. Detecting such deviations has become an active research topic, with approaches ranging from activation fingerprints~\cite{toploc,svip} to trusted hardware~\cite{aftune,veriattn} and replicated execution~\cite{verde}.

\begin{figure}[t]
	\centering
	\includegraphics[width=\columnwidth]{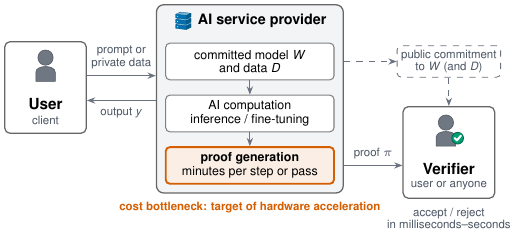}
	\caption{Verifiable AI as a service. The provider attaches a zero-knowledge proof to each result; verification is cheap, while proof generation dominates the cost.}
	\label{fig:overview}
\end{figure}

Zero-knowledge proofs provide cryptographic assurance about a precisely specified computation. As illustrated in Fig.~\ref{fig:overview}, a provider attaches a short proof to its answer, and anyone can check the proof in milliseconds to seconds. A valid proof certifies that the encoded relation holds under the proof system's security assumptions, while hiding the designated private inputs. When that relation binds the model, input and computation, it certifies the requested inference or training task. Unlike trusted hardware, the guarantee rests on cryptographic assumptions rather than on trust in a chip vendor, and the proof can be verified publicly. Systems now exist for proving inference of large language models~\cite{zkllm,zkgpt,deepprove,zkagent}, fine-tuning steps~\cite{verilora,vft}, and differentially private training~\cite{veridp}. The obstacle is cost on the prover side. Proving a single forward pass of a 13-billion-parameter model takes close to 15 minutes~\cite{zkllm}; proving one fine-tuning step with low-rank adaptation for models of 3 to 13 billion parameters takes two to four minutes on a data-center graphics processing unit (GPU), about three orders of magnitude longer than the step itself~\cite{verilora}; and a state-of-the-art prover for whole generated sequences reports 86--174 tokens per minute on a single machine~\cite{deepprove}, illustrating the cost of certifying generation. Verification is cheap; proving is the bottleneck. Reducing this cost requires hardware acceleration together with proof-system and implementation choices.

There is a substantial body of work on accelerating zero-knowledge provers on GPUs, field-programmable gate arrays (FPGAs) and application-specific integrated circuits (ASICs)~\cite{hwsurvey_cje}. Most of it, however, targets individual cryptographic kernels---above all multi-scalar multiplication (MSM) and the number-theoretic transform (NTT)---and reports speed-ups on generic benchmarks~\cite{pipezk,gzkp,cuzk,optimsm,szkp}. A kernel speed-up alone does not establish an end-to-end gain: its value depends on the kernel's share of total proving time and on the costs of feeding it data. Proofs of neural networks combine interactive summation checks (the sumcheck protocol) for matrix multiplications, lookup arguments for non-linear functions, commitments to and openings of billions of fixed weights, and recomputation of every intermediate activation over a finite field~\cite{zkllm,deepprove,verilora}. Evidence from GPU studies points the same way: once one kernel is accelerated, the bottleneck moves to the next one---from MSM to NTT~\cite{zkprophet}, or from sumcheck to the polynomial commitment~\cite{chimera}. Moreover, hardware acceleration work has rarely validated its designs on AI workloads: only three of the twenty representative accelerators reviewed in this article report an evaluation on a machine learning model or circuit~\cite{batchzk,zkflex,legozk}, all on convolutional networks.

This article treats proof generation as a layer of AI infrastructure alongside training clusters, inference serving and data pipelines. We organize the literature with a workload-oriented framework and add original measurements of six provers and of the seven cryptographic kernels they rely on. Our measurements show that (i) no single cost dominates, as the largest share shifts with the proof system and the input length (the number of tokens in the context window); (ii) the cost of opening fixed weights does not grow with the input length; and (iii) memory capacity limits what can be proved before data movement limits how fast. The contributions of this article are as follows.
\begin{itemize}
  \item \textbf{Four-layer framework} (Section~\ref{sec:framework}): We decompose verifiable AI into four layers, from verifiable AI workloads (Layer~1) through proof systems (Layer~2) and cryptographic kernels (Layer~3) to hardware platforms (Layer~4), identify the connections between components of adjacent layers, analyze where the proving cost arises at these connections, and highlight four structural properties of AI workloads that hardware can exploit to reduce it.
  \item \textbf{Cross-system measurements} (Section~\ref{sec:measure}): We measure cost breakdowns for inference, low-rank fine-tuning and small-scale differentially private training, supplemented by block-level comparisons and CPU/GPU kernel micro-benchmarks, and show how the dominant cost changes across implementations and scales and how memory capacity constrains feasible workloads.
  \item \textbf{Workload-aware hardware review} (Section~\ref{sec:hw}): We review GPU, FPGA, ASIC and AI-accelerator designs, assess the fit between each kernel and each platform, and identify where accelerator evaluations fall short of AI workloads.
  \item \textbf{Open problems and roadmap} (Section~\ref{sec:open}): We set out seven open problems for making verifiable AI practical, from proof-friendly numeric formats to commitment on the network data path, and order them by time horizon.
\end{itemize}

Existing surveys provide complementary views of this design space. A recent survey of zk-SNARK hardware accelerators~\cite{hwsurvey_cje} covers generic proof systems without an AI workload perspective or end-to-end measurements. A survey of zero-knowledge proof applications~\cite{zkpappsurvey} compares CPU, GPU, FPGA and ASIC platforms only briefly. A broad survey of zero-knowledge verifiable machine learning and a focused perspective on zero-knowledge federated learning~\cite{zkmlsurvey,zkflsurvey} organize protocols, systems and application roles. This article complements that coverage with cross-system cost measurements and a workload-oriented assessment of hardware acceleration. Systematizations of private inference~\cite{sok_privinf1,sok_privinf2} address confidentiality through homomorphic encryption and multi-party computation, not verifiability. A systematization of hash-based polynomial commitments~\cite{sok_hashpcs} studies their soundness, whereas we discuss the choice of commitment scheme from the hardware and performance side.

\section{Verifiable AI: What Is Proved and Why It Is Expensive}
\label{sec:background}

\subsection{Three Classes of Verifiable AI Tasks}
\label{sec:tasks}

Table~\ref{tab:tasks} summarizes three classes of existing zero-knowledge systems for AI, grouped by what the proof certifies. \emph{Verifiable inference} proves that an output was produced by a committed model on a given input. Systems such as zkLLM and zkGPT prove one forward pass~\cite{zkllm,zkgpt}. Recent systems prove an entire generated sequence at once, avoiding repeated proofs of the growing context for every new token~\cite{deepprove,zkagent}.

\emph{Verifiable training and fine-tuning} proves that each update step---forward pass, backward pass and parameter update---was executed on the committed data and weights. Parameter-efficient fine-tuning is the natural starting point because only a small adapter changes while the large base model stays fixed~\cite{verilora,vft}. Per-step proofs can be folded into a single certificate for a whole training run~\cite{vft}.

\emph{Training audit and privacy proofs} certify a property of training rather than every arithmetic operation. A proof of differentially private stochastic gradient descent shows, step by step, that per-sample gradients were clipped and the declared noise was added~\cite{veridp}. A proof of training can instead certify that a committed checkpoint reaches a claimed objective value on inputs chosen by an auditor, at a cost independent of the number of training steps~\cite{zkllmpot}.

\begin{table*}[t]
\centering
\caption{Three classes of verifiable-AI tasks and representative zero-knowledge systems. All costs are as reported by the authors, on the stated hardware, at different security levels, fields, quantization settings and input lengths, and are \emph{not} directly comparable; values in parentheses are normalized per token or per step (notes below). Section~\ref{sec:measure} reports measured cost shares under a common taxonomy.}
\label{tab:tasks}
\footnotesize
\setlength{\tabcolsep}{4pt}
\renewcommand{\arraystretch}{1.15}
\begin{tabular}{@{}>{\raggedright\arraybackslash}p{1.75cm} >{\raggedright\arraybackslash}p{3.1cm} l >{\raggedright\arraybackslash}p{2.0cm} >{\raggedright\arraybackslash}p{4.2cm} >{\raggedright\arraybackslash}p{2.2cm}@{}}
\toprule
\textbf{Task} & \textbf{What the proof certifies} & \textbf{System} & \textbf{Model / scale} & \textbf{Reported cost (normalized)} & \textbf{Platform} \\
\midrule
Verifiable inference
 & The output was produced by the committed model on the given input (one forward pass or a whole generated sequence)
   & zkLLM~\cite{zkllm}       & up to 13B            & $<$15\,min per 2048-token forward pass ($\approx$0.39\,s/token$^{a}$); proof $<$200\,kB & 1$\times$A100 (40\,GB) \\
 & & zkGPT~\cite{zkgpt}       & GPT-2                & 21.8\,s per inference ($\approx$0.68\,s/token$^{b}$) & CPU server (Xeon 6126) \\
 & & DeepProve~\cite{deepprove} & GPT-2 124M, Gemma~3 270M & 174 / 86 tokens per minute, whole sequence (0.34--0.47 / 0.69--0.94\,s/token$^{c}$); verify 1--3.7\,s & CPU servers (EPYC 9254; Ryzen~9) \\
 & & zkAgent~\cite{zkagent}   & GPT-2, 512 tokens    & $\approx$0.30\,s per generated token$^{d}$ & CPU server (Xeon 6982P-C) \\
 & & Attestable$^{\ast}$~\cite{attestable} & Gemma~4 31B & 53 tokens/s, batch 1 (0.019\,s/token); proof 4.4--7.9\,MiB & 1$\times$H100 \\
\midrule
Verifiable training / fine-tuning
 & Each update step (forward, backward, update) ran on the committed data and weights
   & VeriLoRA~\cite{verilora} & 3B--13B (LoRA) & 122--249\,s per step ($\approx$10$^3\times$ plain step); commitment 156--554\,s & 1$\times$A100 (80\,GB) \\
 & & VFT~\cite{vft}           & n/r (LoRA $r$=8/16, 2048 tokens) & 16.8--31.2\,s per step; verify $<$200\,ms & 2--4$\times$80\,GB GPUs \\
\midrule
Training audit and DP proofs
 & Each DP-SGD step clipped per-sample gradients and added the declared noise
   & VeriDP~\cite{veridp}     & up to 11.2M params   & 0.34--8.5\,s per sample; memory is the main limit & 1 CPU core \\
\cmidrule(l){2-6}
 & A committed checkpoint attains the claimed objective on auditor-chosen inputs
   & zkLLMPoT~\cite{zkllmpot} & 1.1B--13B            & covered operators of one 512-token forward audit: 41--59\,s (1.1--1.5B), 131\,s (13B) (0.08--0.26\,s/token); verify $<$0.5\,s & 1$\times$A100 (40\,GB) \\
\bottomrule
\end{tabular}

\par\vspace{2pt}
\raggedright\scriptsize
$^{\ast}$Vendor blog post, not peer-reviewed. n/r: not reported.
$^{a}$803\,s per 2048-token prompt pass of the 13B model.
$^{b}$Input length inferred from matrix dimensions (32$\times$768), not stated.
$^{c}$Peak tokens per minute (TPM) of GPT-2 / Gemma~3 on the faster of two reported machines; per token = 60/TPM over both machines.
$^{d}$v3 abstract; the earlier v2 paper reports 0.40\,s/token for the same setting.
\end{table*}

\subsection{Zero-Knowledge Proofs in a Nutshell}
\label{sec:zkp}

\begin{figure}[t]
\centering
\includegraphics[width=\columnwidth]{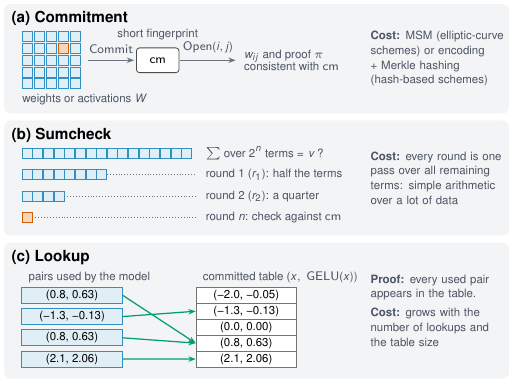}
\caption{The three building blocks that dominate proofs of neural networks.}
\label{fig:zkp}
\end{figure}

To prove a computation, it is first written as a set of algebraic constraints over a finite field---for example, integers modulo a prime. A matrix multiplication becomes many multiply-and-add constraints; a softmax becomes a combination of arithmetic constraints and table lookups. The prover then convinces the verifier that it knows values for all intermediate variables (the \emph{witness}) that satisfy every constraint, without revealing them. Fig.~\ref{fig:zkp} illustrates three building blocks that recur in these proofs.

\textbf{Commitment.} A commitment locks a large piece of data---a weight matrix, a layer's activations---into a short fingerprint. Afterwards the prover cannot change the data, yet it can later \emph{open} the commitment at chosen positions and prove that the revealed values are consistent with the fingerprint. Hash-based commitments build a Merkle tree over an encoded version of the data~\cite{fri,brakedown,basefold}; elliptic-curve-based commitments compute the fingerprint as a weighted sum of millions of curve points, an operation called multi-scalar multiplication (MSM)~\cite{kzg,hyrax}.

\textbf{Sumcheck.} Many claims about a computation take the form ``this sum over a huge number of terms equals $v$''. An entry of a matrix product is exactly such a sum. The sumcheck protocol~\cite{lfkn} reduces a sum over $2^n$ terms to $n$ short rounds; in each round the verifier sends a random challenge and the prover answers with a low-degree polynomial, so a cheating prover is caught with overwhelming probability. For the prover, each round is a pass over all remaining terms that halves the problem size---simple arithmetic, but over a lot of data. GKR-style systems chain sumchecks layer by layer through a whole network~\cite{gkr}.

\textbf{Lookup.} Non-linear functions such as GELU, softmax exponentials or layer-norm square roots are expensive to express as arithmetic. Instead, all valid input--output pairs are placed in a pre-committed table, and a lookup argument proves that every pair used in the computation appears in the table~\cite{logup,zkllm}. The cost grows with the number of lookups and the size of the table. Two further kernels appear depending on the proof system. The number-theoretic transform (NTT) is the finite-field analog of the fast Fourier transform and converts between polynomial representations in univariate systems such as Plonk~\cite{plonk}. Hashing---with arithmetic-friendly functions such as Poseidon2~\cite{poseidon2} or bit-oriented ones such as BLAKE3~\cite{blake3}---builds the Merkle trees used by hash-based commitments.

\subsection{Why Proving AI Is Expensive}
\label{sec:why}

Four factors combine to make the prover several orders of magnitude slower than the computation it proves.

\textbf{Finite-field arithmetic instead of low-precision floating point.} AI hardware is built for 16-bit and 8-bit floating-point or integer matrix multiplication. Proofs operate on field elements that are either about 256 bits wide (for elliptic-curve-based systems) or 31 to 64 bits wide (for recent small-field systems~\cite{binius}). A single 256-bit modular multiplication requires multiple lower-width integer operations, and GPU provers run almost entirely on the integer pipeline~\cite{zkprophet}; the tensor cores that power AI were left unused until very recently~\cite{tensorzkp}. In the measured provers, models are quantized to integer or fixed-point representations before their computations are expressed over a field.

\textbf{The witness is much larger than the model.} A proof must account for every intermediate value: all activations, every lookup query and its multiplicity, and, for training, all gradients. The witness of a fine-tuning step therefore exceeds the model size by a large factor: proving one low-rank fine-tuning step of a 7-billion-parameter model allocates about 2.9\,TB of GPU memory in total across layers, which fits on a single 80\,GB GPU only because layers are processed one at a time~\cite{verilora}.

\textbf{Commitments touch all of this data.} Every committed tensor must be encoded and hashed, or multiplied into an MSM. For hash-based schemes, committing to and proving a single polynomial with $2^{28}$ coefficients already involves about 1.1\,GB of data and takes 463\,s on a high-end server CPU~\cite{hashpcs_bw}.

\textbf{Data movement.} At larger model sizes or input lengths, the witness can exceed accelerator memory. It is generated, moved between host and device, committed, and later reloaded to answer opening queries, so memory capacity, and with it the transfers needed to work around it, become first-order constraints alongside arithmetic~\cite{batchzk,zkpog}.

\subsection{Other Routes to Verifiability}
\label{sec:other}

Zero-knowledge proofs are not the only way to check an AI service. Trusted execution environments~(TEE)~\cite{aftune,veriattn} are fast but require trusting the hardware vendor and the enclave implementation. Replicated execution with refereed disputes~\cite{verde}, made reproducible by removing hardware non-determinism~\cite{optimistic}, needs only one honest provider among several but is not publicly verifiable by a single party. Activation fingerprints~\cite{toploc}, learned proxy tasks~\cite{svip} and checkpoint replay in proof-of-learning~\cite{pol} are cheap but give statistical rather than cryptographic guarantees. This article focuses on cryptographic proofs because they combine public verifiability with independence from hardware trust---and because their cost is the problem that hardware can most directly address.

\section{A Four-Layer Framework from Verifiable AI Workloads to Hardware}
\label{sec:framework}

\begin{figure*}[t]
\centering
\includegraphics[width=\textwidth]{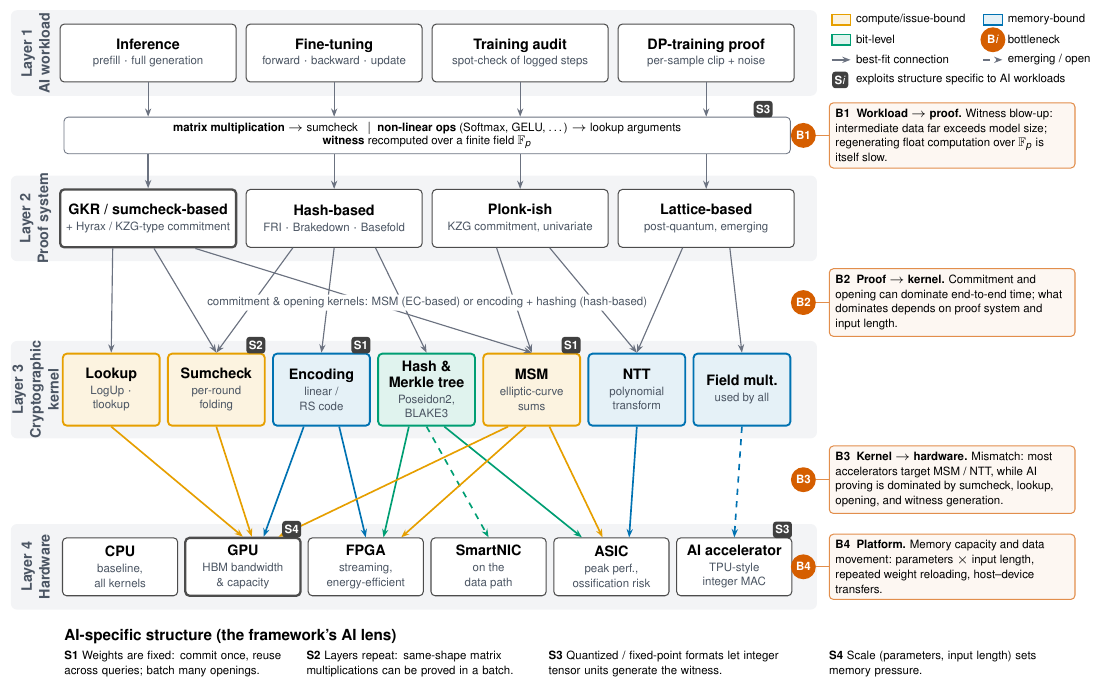}
\caption{Four-layer framework. Each verifiable AI workload (Layer~1) is expressed in a proof system (Layer~2), whose cost is concentrated in a few cryptographic kernels (Layer~3) that run with different efficiency on hardware platforms (Layer~4). Kernel colors give the limiting resource as measured on a GPU (Section~\ref{sec:kernels}); B1--B4 mark the bottleneck introduced at each connection between layers; S1--S4 mark where structure specific to AI workloads can be exploited.}
\label{fig:framework}
\end{figure*}

Hardware choices for verifiable AI are hard to reason about because the cost of a proof is determined by a chain of decisions: which workload is proved, which proof system expresses it, which kernels that proof system spends its time in, and which hardware runs those kernels well. We organize these decisions into four layers and examine the connections between adjacent layers, each of which introduces a bottleneck. Fig.~\ref{fig:framework} shows the four layers, the connections between them and the bottleneck introduced at each connection.

\subsection{The Four Layers}
\label{sec:layers}

\textbf{Layer 1: verifiable AI workloads.} A \emph{workload} is the computation that the prover must process to certify a verifiable AI task of Section~\ref{sec:tasks}. We distinguish four kinds: inference (a single prefill or a whole generated sequence), fine-tuning (forward, backward and update), training audit, and differential-privacy proofs. They differ in size, in how much of the computation repeats, and in which parts of the model stay fixed.

\textbf{Layer 2: proof systems.} We distinguish four families. \emph{GKR/sumcheck-based} systems prove layered computations through chained sumchecks and are paired with a separate polynomial commitment such as Hyrax or a KZG variant~\cite{gkr,hyrax,kzg}; most large-model provers belong here~\cite{zkllm,verilora,deepprove}. \emph{Plonkish} systems compile the computation into a circuit with custom gates; the original design relies on univariate polynomials and the NTT~\cite{plonk}, while its multilinear successor replaces the NTT with sumcheck~\cite{hyperplonk}. \emph{Hash-based} systems use error-correcting codes and Merkle trees for commitments, avoid elliptic curves, and are plausibly post-quantum~\cite{fri,brakedown,basefold,binius}. \emph{Lattice-based} systems are an emerging post-quantum alternative~\cite{labrador}. The interactive part of a proof and the commitment scheme are largely independent choices, which is why the same sumcheck-based prover can be paired with very different commitments.

\textbf{Layer 3: cryptographic kernels.} Seven kernels account for most prover time: MSM, NTT, sumcheck, lookup, hashing with Merkle-tree construction, encoding with linear or Reed--Solomon codes, and the underlying finite-field multiplication.

\textbf{Layer 4: hardware platforms.} CPUs serve as the baseline. GPUs are the workhorse of current provers. FPGAs offer customizable streaming pipelines, ASICs specialized datapaths, AI accelerators such as TPUs massive integer matrix throughput, and smart network interface cards (SmartNICs) computation on the data path.

\subsection{Connections Between Layers and Their Bottlenecks}
\label{sec:connections}

\textbf{From workloads to proof systems (B1).} The structure of neural networks steers the choice of proof system. Matrix multiplication, which dominates both inference and training, is a sum of products and can be proved directly with sumcheck, at a prover cost linear in the matrix size~\cite{zkllm}. Non-linear operations are proved with lookup arguments. The price is the witness of Section~\ref{sec:why}, recomputed over the field and later committed, so the first bottleneck is \emph{witness blow-up and field recomputation}, which exists before any cryptographic kernel runs.

\textbf{From proof systems to kernels (B2).} Each proof system concentrates its cost in a few kernels. Elliptic-curve commitments spend their time in MSM; hash-based commitments in encoding and hashing; Plonk-style provers additionally in NTT; sumcheck and lookup arguments in passes over large tables. For AI workloads the commitment side can dominate end-to-end time, because the witness is large and the weights must be opened again and again, and once sumcheck is accelerated on a GPU the commitment becomes the next bottleneck~\cite{chimera}. Whether it dominates in practice depends on the proof system and the input length, as Section~\ref{sec:breakdown} shows. The second bottleneck is \emph{commitment and opening}.

\textbf{From kernels to hardware (B3, B4).} Kernels differ in the resource that limits them. MSM consists of many independent elliptic-curve additions with high arithmetic intensity and is compute-bound; it parallelizes well on GPUs and ASICs~\cite{gzkp,szkp}. Sumcheck, lookup and encoding stream over large tables with little data reuse, which the literature treats as bandwidth-bound; whether sumcheck or NTT is the better primitive depends on polynomial degree and on-chip memory~\cite{nttsumcheck}. Hashing consists of permutation rounds and bitwise operations and suits deep streaming pipelines. Our micro-benchmarks (Section~\ref{sec:kernels}) refine this picture: on a GPU, encoding and element-wise field multiplication are indeed memory-bound, but low-degree sumcheck reaches the integer-instruction issue limit, while lookup is limited by synchronization and instruction issue. Table~\ref{tab:kernel-hw} summarizes the resulting fit, with the limiting resource taken from those measurements. The third bottleneck is a \emph{mismatch}: most accelerators were designed for MSM and NTT, while AI provers also incur substantial lookup, sumcheck and witness-generation costs, as measured in Section~\ref{sec:breakdown}. The fourth is the platform itself: \emph{memory capacity and data movement}, because model parameters, input length and repeated weight reloading together exceed what a single device holds.

\providecommand{\fitS}{\CIRCLE}
\providecommand{\fitM}{\LEFTcircle}
\providecommand{\fitW}{\Circle}
\providecommand{\fitN}{--}
\providecommand{\unv}{\textsuperscript{?}}
\begin{table*}[t]
\centering
\caption{Fit between cryptographic kernels and hardware platforms. Ratings are qualitative and cite representative works. The \emph{Bound by} column gives the resource that limits each kernel on a GPU as measured in Section~\ref{sec:kernels} (``int.\ issue'': integer-instruction issue); Fig.~\ref{fig:kernels} shows measured throughput for the CPU and GPU columns.}
\label{tab:kernel-hw}
\footnotesize
\setlength{\tabcolsep}{3.5pt}
\renewcommand{\arraystretch}{1.2}
\begin{tabular}{@{}l l c c c c c c >{\raggedright\arraybackslash}p{3.6cm}@{}}
\toprule
\textbf{Kernel} & \textbf{Bound by} & \textbf{CPU} & \textbf{GPU} & \textbf{FPGA} & \textbf{ASIC} & \textbf{AI acc.} & \textbf{SmartNIC} & \textbf{Main reason} \\
\midrule
MSM & compute
  & \fitW & \fitS~\cite{gzkp,cuzk,icicle} & \fitM~\cite{optimsm,ifzkp,legozk} & \fitS~\cite{pipezk,szkp} & \fitM~\cite{aiasic} & \fitN
  & Many independent elliptic-curve additions; high arithmetic intensity, parallelizes well. \\
NTT & memory
  & \fitW & \fitS~\cite{gzkp,zkprophet} & \fitM~\cite{legozk} & \fitS~\cite{pipezk,szkp} & \fitS~\cite{aiasic} & \fitN
  & Butterfly stages with strided access; limited by memory-issue queues and occupancy rather than DRAM bandwidth or integer arithmetic. \\
Sumcheck & int.\ issue
  & \fitM~\cite{chimera} & \fitS~\cite{batchzk,chimera,tensorzkp} & \fitM~\cite{pipesc} & \fitM~\cite{zkspeed,zkphire,pipesc,mtu} & \fitN & \fitN
  & Each round streams over all evaluations with little reuse; low-degree rounds saturate integer issue before bandwidth, and the best primitive depends on degree and on-chip memory~\cite{nttsumcheck}. \\
Lookup & sync., issue
  & \fitM & \fitS~\cite{zkllm} & \fitN & \fitN & \fitN & \fitN
  & Random table accesses plus batched field inversions (LogUp~\cite{logup}); barriers in the inversion, not DRAM, cap the GPU gain. \\
Hash \& Merkle & bit-level
  & \fitM & \fitS~\cite{batchzk} & \fitS~\cite{hashpcs_bw} & \fitS~\cite{nocap,mtu} & \fitW & open
  & Permutation rounds and bitwise operations, limited by instruction issue rather than DRAM; tree reduction suits streaming pipelines. \\
Encoding & memory
  & \fitM & \fitS~\cite{batchzk,tensorzkp} & \fitM~\cite{hashpcs_bw} & \fitS~\cite{nocap} & \fitM~\cite{aiasic} & \fitN
  & Sparse linear codes or Reed--Solomon (via NTT); off-chip bandwidth limits the gain~\cite{hashpcs_bw}. \\
Field mult. & memory
  & \fitM & \fitS & \fitM & \fitS & \fitM~\cite{aiasic} & \fitW
  & 256-bit fields need multi-limb arithmetic, small fields (31/64-bit) map to native integer units; element-wise, it is memory-bound. \\
\bottomrule
\end{tabular}

\vspace{2pt}
\raggedright\scriptsize
\fitS~strong \quad \fitM~medium \quad \fitW~weak \quad --~no published evidence. Tensor cores inside GPUs~\cite{tensorzkp} are counted under GPU; ``AI acc.'' means stand-alone AI accelerators such as TPUs~\cite{aiasic}. ``open'': SmartNICs have been proposed for observing training traffic~\cite{frontierzk}, but no published design hashes or commits tensors on the data path as discussed in Section~\ref{sec:open}.
\end{table*}

\subsection{Structure Specific to AI Workloads}
\label{sec:ai-structure}

AI workloads expose regularities beyond those represented by isolated circuit benchmarks. Fig.~\ref{fig:framework} summarizes four such properties (S1--S4), which provide opportunities for reuse, batching and coordinated memory management.

\textbf{S1: weights are fixed.} The base model does not change across inference requests, and during parameter-efficient fine-tuning it does not change across steps. Its commitment can be computed once and reused---existing provers already separate a one-off weight commitment from per-query proving~\cite{zkllm,deepprove}---and the many openings that each request or step requires can be merged into fewer, larger ones~\cite{shplonk,zkvalue}. This shifts the cost from repeated commitment to efficient opening.

\textbf{S2: layers repeat.} A transformer applies the same matrix shapes in every layer and for every request. Proofs of same-shape operations can be batched and pipelined to raise hardware utilization, an approach that already yields large throughput gains on GPUs~\cite{batchzk}.

\textbf{S3: numeric formats are a design choice.} If a model is quantized to integers or fixed-point numbers that the proof system can absorb, for example because every intermediate result, including accumulated sums of products, stays below the field modulus, the witness can be produced by the same integer matrix units that run the model, rather than by slow field arithmetic. Tensor cores~\cite{tensorzkp} and TPU matrix units~\cite{aiasic} have already been repurposed for finite-field kernels, and an industrial prover reports proving 8-bit integer matrix multiplication at large scale~\cite{attestable}.

\textbf{S4: scale sets memory pressure.} Parameter count and input length determine the witness size and therefore whether a proof fits on one device. Because both are known before proving starts, the memory a workload needs can be predicted and the hardware provisioned accordingly. Proving a whole generated sequence at once~\cite{deepprove,zkagent} trades redundant per-token work for a larger single proof, which makes memory capacity and bandwidth even more decisive.

Used this way, the framework turns ``which hardware should accelerate verifiable AI?'' into a sequence of answerable questions: for a given workload, which proof system is used, which kernels dominate, which platforms suit them, and which AI-specific structure can reduce the work before any hardware is involved. Section~\ref{sec:measure} follows this procedure with measurements on four workloads, and Section~\ref{sec:hw} uses it to assess existing accelerators.

\section{Where the Cost Goes: Measurements}
\label{sec:measure}

The framework of Section~\ref{sec:framework} predicts where proving time should go for AI workloads. This section tests those predictions. We measure a set of workloads across publicly available provers and use their cost shares and scaling trends to identify recurring bottlenecks. Implementation differences remain part of the comparison and are accounted for when interpreting the results.

\subsection{Methodology}
\label{sec:method}

\textbf{Questions.} The measurements address four questions derived from the framework:
\begin{itemize}
  \item[Q1] Do commitment and opening dominate proving time across proof systems (B2)?
  \item[Q2] What share of the time goes to generating the witness over the finite field (B1)?
  \item[Q3] How do the cost components scale with model size and input length, and is the cost of opening fixed weights independent of input length (S1, S4)?
  \item[Q4] When do memory capacity and data movement become the binding constraint (B4)?
\end{itemize}

\textbf{Workloads.} (W1) Inference over a prompt: a 7-billion-parameter LLaMA-2 model with prompts of 1024 and 2048 tokens on zkLLM (shorter prompts are structurally unsupported by its code), and a 0.27-billion-parameter Gemma~3 model with prompts of 64--1024 tokens on DeepProve, which did not run the 7-billion-parameter model in any configuration we tried. (W2) Proving a whole generated sequence, on DeepProve only. (W3) One fine-tuning step with low-rank adaptation for 7- and 13-billion-parameter models at input lengths 64--2048. (W4) One step of differentially private stochastic gradient descent, at the largest scale its prover supports, far below that of W1--W3.

\textbf{Provers.} Table~\ref{tab:systems} lists the provers: public code, end-to-end support for the workload, and as wide a coverage of Layer~2 as possible. Most open-source provers that reach the scale of large language models are built on sumcheck and GKR~\cite{zkllm,deepprove,verilora}; we found no hash-based prover that runs one end to end. We therefore add two narrower points. A GKR prover with interchangeable commitment schemes~\cite{expander} runs the same transformer block with an elliptic-curve commitment (Hyrax or KZG) and a hash-based one (Orion), and a compiler built on KZG blocks and proof accumulation~\cite{zktorch} provides a point outside the sumcheck family. All provers are used at fixed commits (Table~\ref{tab:systems}), with timing code added where a prover does not report its own breakdown and no change to what is proved. DeepProve~\cite{deepprove} is run unmodified: its built-in timers already report the per-stage times that we use for the breakdown.

\begin{table}[t]
\centering
\caption{Provers used in the measurements.}
\label{tab:systems}
\scriptsize
\setlength{\tabcolsep}{2.5pt}
\renewcommand{\arraystretch}{1.12}
\begin{tabular}{@{}l >{\raggedright\arraybackslash}p{2.35cm} l >{\raggedright\arraybackslash}p{1.45cm} l@{}}
\toprule
\textbf{Prover} & \textbf{Proof system / commitment} & \textbf{Workload} & \textbf{Role} & \textbf{Commit} \\
\midrule
zkLLM~\cite{zkllm}         & sumcheck, lookup; Hyrax-style (BLS12-381) & W1 & full scale & \texttt{993311e} \\
DeepProve~\cite{deepprove} & GKR, lookup; HyperKZG (BN254)             & W1, W2 & small model & \texttt{9d1a53e} \\
VeriLoRA~\cite{verilora}   & sumcheck, lookup; Hyrax-style (BLS12-381) & W3 & full scale & \texttt{7cc0977} \\
VeriDP~\cite{veridp}       & GKR; incremental verification             & W4 & small scale & \texttt{cf86a46} \\
Expander~\cite{expander}   & GKR; Orion / Hyrax / KZG                   & one block & commitment comparison & \texttt{096581e} \\
ZKTorch~\cite{zktorch}     & KZG blocks, accumulation (BN254; CPU)      & MLP block & non-sumcheck point & \texttt{63b9c68} \\
\bottomrule
\end{tabular}
\par\vspace{2pt}
\raggedright\scriptsize Kernel libraries: ICICLE v2.8.0 (GPU) and v4.0.0 (CPU), sppark \texttt{69440fd}.
\end{table}

\textbf{Cost taxonomy.} To compare provers that report time in different ways, we map every timed region to one of eight categories aligned with the framework: \emph{witness generation} (recomputing the model over the field, including quantization; B1), \emph{sumcheck}, \emph{lookup}, \emph{commitment}, \emph{opening} (proving evaluations of committed data; B2), \emph{hashing and encoding}, \emph{data movement} (host--device and disk transfers; B4), and \emph{other}. Regions that cannot be split without modifying the code are reported as \emph{mixed} rather than apportioned by guesswork. One-off setup---key generation and the commitment to fixed weights---is reported separately from per-query cost, because amortizing it is exactly the opportunity S1.

\textbf{Metrics and kernels.} For each cell we record prover time and its breakdown, peak GPU and host memory, proof size, verifier time, and the slowdown relative to the same computation in 16-bit floating point with PyTorch on the same GPU. Because the provers differ in field, security level and quantization, we compare \emph{cost shares} and \emph{scaling trends}, not absolute times. To connect the breakdowns to hardware, we also benchmark the seven kernels of Layer~3 in isolation, using GPU libraries~\cite{icicle,sppark} or reference implementations, and place each on a roofline built from measured rather than vendor-specified roofs.

\textbf{Platforms and protocol.} A shared server (five NVIDIA A100-PCIE-40GB cards; dual Xeon Gold 6248R, 96 logical cores) ran the kernel micro-benchmarks, the zkLLM breakdowns and the CPU-only provers; an exclusively used server (one A100-PCIE-80GB card of the same architecture; 192-core host) ran VeriLoRA, DeepProve, the commitment comparison and the engineering waterfall. Absolute times are not compared across the two. Each configuration is run once to warm up and three times for measurement, and we report the median; on the shared server every repetition is gated on low system load. A few long configurations of DeepProve (up to 22 minutes each) have a single repetition, as Table~\ref{tab:overheads} states. FPGA and ASIC results are taken from the literature and are not mixed with our measurements.

\textbf{Limitations.} Full-scale coverage is mostly restricted to sumcheck-based provers; the non-sumcheck point is a CPU-only multilayer-perceptron block, and the commitment comparison is made on one transformer block. The two inference provers differ in commitment scheme and implementation, so differences between them have no single cause, and the second one reaches only 0.27 billion parameters; its author-reported rates are CPU-only, so most of its cells give a breakdown rather than a rate comparison. W4 runs on a single-threaded CPU prover with small networks and indicates trends only. Capacity (Q4) is observed at two points, 40 and 80\,GB, not as a continuous comparison.

\subsection{Kernel Throughput Across Platforms}
\label{sec:kernels}

\begin{figure*}[t]
\centering
\includegraphics[width=\textwidth]{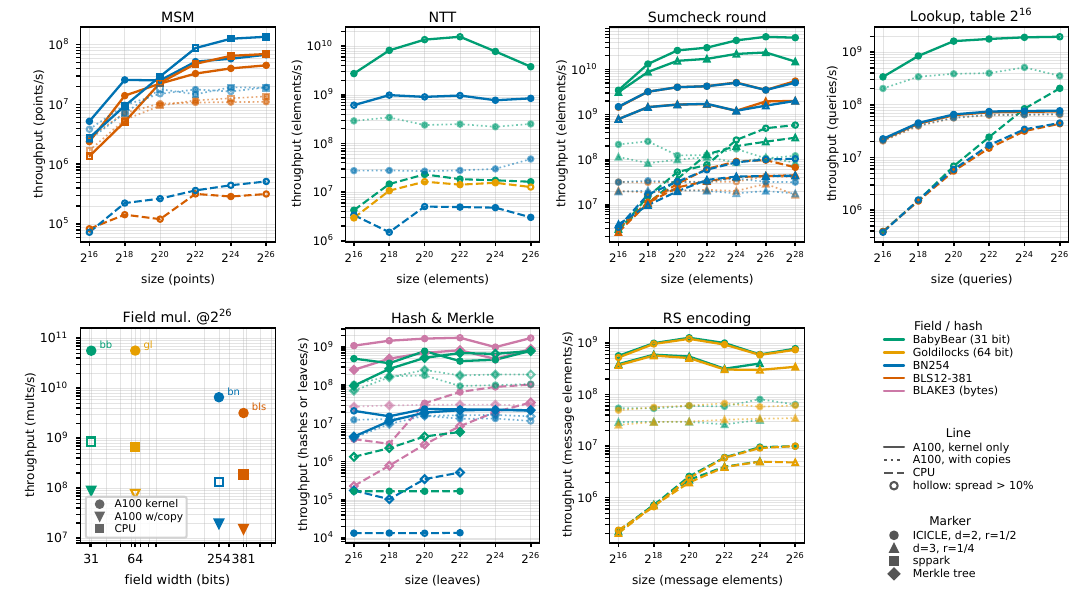}
\caption{Throughput of the seven Layer-3 kernels (log scale) at $2^{16}$--$2^{26}$ elements on the A100 GPU and the host CPU. Solid lines: GPU execution time with data resident on device; dotted: including host--device copies; dashed: CPU. Color gives the field (or hash); markers separate sumcheck degrees ($d$), Reed--Solomon rates ($r$), the two MSM libraries and hashing from Merkle-tree construction. Hollow markers denote runs whose repetition spread exceeded 10\%. No GPU NTT is shown for Goldilocks because the publicly usable library release lacks that field.}
\label{fig:kernels}
\end{figure*}

\begin{figure}[!b]
\centering
\includegraphics[width=\columnwidth]{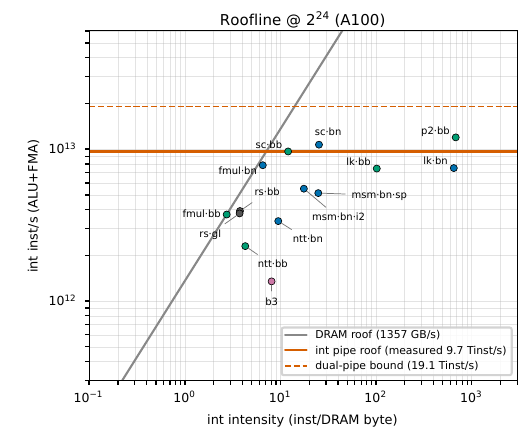}
\caption{Roofline of the kernels at $2^{24}$ elements on the A100: thread-level integer-instruction rate, summed over the ALU and FMA pipelines, against integer instructions per DRAM byte. The diagonal is the measured copy bandwidth ($\approx$1.36\,TB/s); the solid and dashed lines are the \emph{measured} single-pipe ($\approx$9.66$\times$10$^{12}$ inst/s) and dual-pipe ($\approx$1.91$\times$10$^{13}$) issue limits, so points may lie above the solid line. Left of the ridge point ($\approx$7 inst/byte) bandwidth can limit a kernel, right of it instruction issue; a point far below both is limited by dependencies, synchronization or occupancy. Labels: sc sumcheck, lk lookup, fmul field mult., rs Reed--Solomon, p2 Poseidon2, b3 BLAKE3; bb BabyBear, gl Goldilocks, bn BN254; i2 ICICLE v2, sp sppark.}
\label{fig:roofline}
\end{figure}

Fig.~\ref{fig:kernels} plots the throughput of the seven kernels, and Fig.~\ref{fig:roofline} places them on a roofline built from the measured roofs of the 40\,GB A100. MSM and NTT use version 2.8 of ICICLE~\cite{icicle}, cross-checked on MSM against a second implementation~\cite{sppark}. CPU MSM and Poseidon2 are single-threaded reference paths, so their GPU/CPU ratios indicate direction only.

\textbf{Compute and bandwidth limits.} MSM is compute-bound in the measured implementation: its dominant GPU kernels reach only 15--23\% of the copy roof and stall on dependencies between large-integer operations. Reed--Solomon encoding is bandwidth-bound, at 74--76\% of the copy roof.

\textbf{Instruction issue and synchronization.} Three kernel classifications that the literature suggests do not hold on this hardware, and Table~\ref{tab:kernel-hw} uses the measured ones. \emph{Sumcheck} (degree~2) is not memory-bound: it saturates the integer-issue roof (100--111\% of the single-pipe roof) while using 31--59\% of the copy roof. \emph{Lookup} is synchronization- and issue-bound: DRAM use is 0.8--5\% of the copy roof, integer issue reaches 77--78\%, and warps stall on barriers inside the batched field inversion. \emph{Field multiplication}, measured element-wise, is memory-bound in both a 31-bit field (at the copy roof) and a 254-bit field (88\%), so its ``compute'' label reflects arithmetic complexity, not measured behavior.

\textbf{Memory access and instruction issue.} NTT stays memory-bound, but it reaches only 26--39\% of the copy roof and 9--17\% of the integer roof; profiling attributes the gap to memory-issue queues (strided access, bank conflicts) and low occupancy, not to the 32-bit integer pipeline that a previous characterization blamed~\cite{zkprophet}. The streaming NTT stage inside our Reed--Solomon encoder reaches 74--76\% of the copy roof on the same card, so data layout decides how close NTT gets to DRAM bandwidth. Hashing is issue-bound in two ways: Poseidon2 on the integer pipes (63\% of the dual-pipe roof), BLAKE3 on load issue (12\% of the copy roof).

\textbf{GPU versus CPU, and the cost of copies.} The GPU/CPU throughput ratio spans three orders of magnitude, from 1.7$\times$ for large-field lookup (capped by the barriers above) to about 2500$\times$ for Poseidon2. Counting host--device copies costs the streaming kernels most---field multiplication 210--741$\times$, sumcheck 113--412$\times$, NTT 15--18$\times$---while MSM loses only 3.5--7$\times$: the kernels that move the most data are the ones for which offloading costs the most, an early sign of the B4 bottleneck.

\subsection{End-to-End Cost Breakdowns}
\label{sec:breakdown}

\begin{figure*}[t]
\centering
\includegraphics[width=\textwidth]{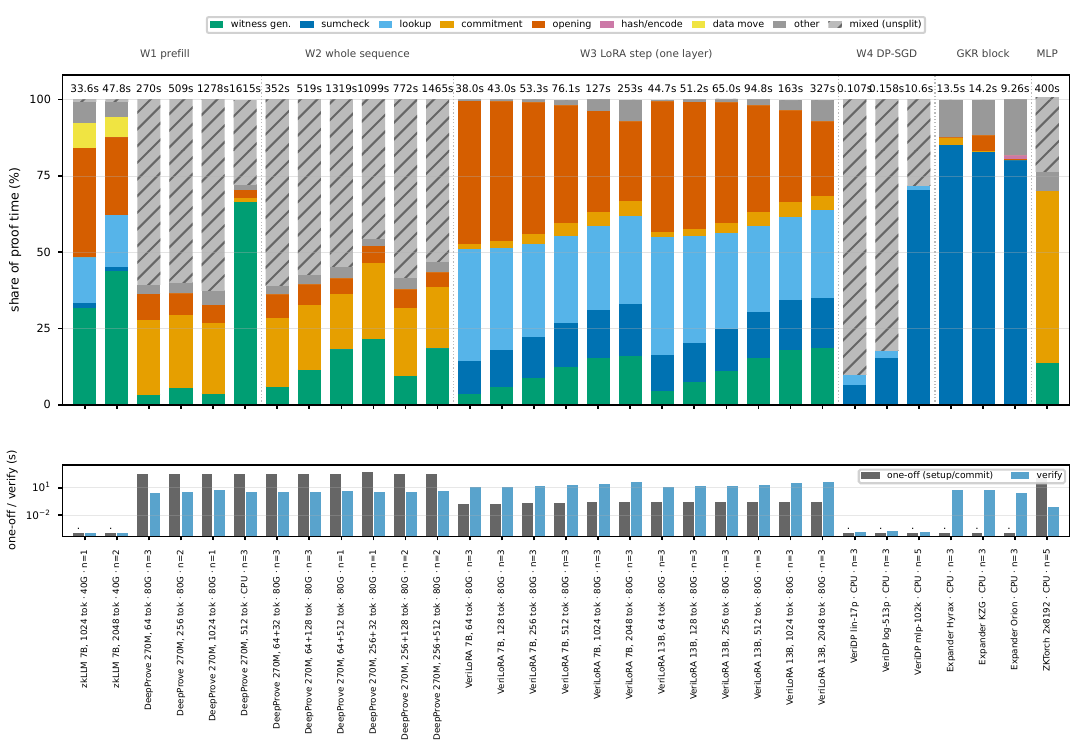}
\caption{Share of proof time by cost category for each measured (workload, prover, scale) cell; bar tops give the median proof time and the lower strip shows one-off setup and verification on a logarithmic scale. Labels give the prover, the model and the input length in tokens (P+G: P prompt plus G generated tokens), the platform (40G: shared A100-40GB; 80G: exclusive A100-80GB; CPU) and the number of repetitions $n$.}
\label{fig:breakdown}
\end{figure*}

\begin{table*}[t]
\centering
\caption{Measured prover overhead per workload, system, and scale. ``prove'' is the median sum of the eight taxonomy categories with one-off setup and verification excluded. Fig.~\ref{fig:breakdown} shows the decomposition. $\times$plain is the ratio over the unmodified PyTorch step measured on the same machine; n/a marks cells with no compatible baseline. A representative subset of cells is shown here; Fig.~\ref{fig:breakdown} shows the full set. The \emph{reps} column gives the number of repetitions $n$ (CPU: host-CPU-only runs). $^{\dagger}$DeepProve times trace generation (executing the quantized model to build the witness) separately, before its proving timer; to keep this column comparable across provers we report end-to-end prover time, and the superscript gives the share of it spent in trace generation.}
\label{tab:overheads}
\footnotesize
\setlength{\tabcolsep}{4pt}
\renewcommand{\arraystretch}{1.15}
\begin{tabular}{@{}l l l r r r r r r@{}}
\toprule
\textbf{Workload} & \textbf{System / scale} & \textbf{reps} & \textbf{prove (s)} & \textbf{verify (s)} & \textbf{proof} & \textbf{GPU pk (GB)} & \textbf{host pk (GB)} & \textbf{$\times$plain} \\
\midrule
  W1 & zkLLM 7B, 1024 tok & n=1 & 33.6 & 1.8e-05 & n/a & 21.2 & 0.84 & 341$\times$ \\
  W1 & zkLLM 7B, 2048 tok & n=2 & 47.8 & 1.5e-05 & n/a & 29.3 & 0.84 & 249$\times$ \\
  W1 & DeepProve 270M, 64 tok & n=3$^{\dagger}$3\% & 270 & 2.8 & 12.66 MB & 21.7 & 60.2 & n/a \\
  W1 & DeepProve 270M, 256 tok & n=2$^{\dagger}$6\% & 509 & 3.28 & 15.29 MB & 65.6 & 95.4 & n/a \\
  W1 & DeepProve 270M, 1024 tok & n=1$^{\dagger}$4\% & 1278 & 5.12 & 18.13 MB & 24.4 & 260.4 & n/a \\
  W1 & DeepProve 270M, 512 tok & n=3, CPU$^{\dagger}$67\% & 1615 & 3.8 & 16.66 MB & -- & 146.4 & n/a \\
\midrule
  W3 & VeriLoRA 7B, 64 tok & n=3 & 38.0 & 11.6 & n/a & 13.6 & 0.14 & 5.0k$\times$ \\
  W3 & VeriLoRA 7B, 512 tok & n=3 & 76.1 & 17.7 & n/a & 18.5 & 0.16 & 9.0k$\times$ \\
  W3 & VeriLoRA 7B, 2048 tok & n=3 & 253 & 39.7 & n/a & 39.9 & 0.52 & 11.1k$\times$ \\
  W3 & VeriLoRA 13B, 64 tok & n=3 & 44.7 & 12.8 & n/a & 20.6 & 0.14 & n/a \\
  W3 & VeriLoRA 13B, 512 tok & n=3 & 94.8 & 20.2 & n/a & 27.2 & 0.16 & n/a \\
  W3 & VeriLoRA 13B, 2048 tok & n=3 & 327 & 46.9 & n/a & 56.2 & 0.52 & n/a \\
\midrule
  W4 & VeriDP lin-17p & n=3, CPU & 0.107 & 1.5e-04 & 3.5 kB & -- & 0.15 & 92$\times$ \\
  W4 & VeriDP log-513p & n=3, CPU & 0.158 & 2.0e-04 & 3.7 kB & -- & 0.15 & 143$\times$ \\
  W4 & VeriDP mlp-102k & n=5, CPU & 10.6 & 1.5e-04 & 3.9 kB & -- & 0.38 & n/a \\
\midrule
  block & Expander Hyrax & n=3, CPU & 13.5 & 5.76 & 939.0 kB & -- & 5.80 & n/a \\
  block & Expander KZG & n=3, CPU & 14.2 & 6.1 & 876.8 kB & -- & 5.80 & n/a \\
  block & Expander Orion & n=3, CPU & 9.26 & 2.4 & 6.25 MB & -- & 5.92 & n/a \\
\midrule
  MLP & ZKTorch 2x8192 & n=5, CPU & 400 & 0.0693 & 14.9 kB & -- & 47.8 & n/a \\
\midrule
  W2 & DeepProve 270M, 64+32 tok & n=3$^{\dagger}$6\% & 352 & 3.06 & 13.95 MB & 14.7 & 71.9 & n/a \\
  W2 & DeepProve 270M, 64+128 tok & n=3$^{\dagger}$12\% & 519 & 3.31 & 15.25 MB & 14.7 & 95.5 & n/a \\
  W2 & DeepProve 270M, 64+512 tok & n=1$^{\dagger}$18\% & 1319 & 3.9 & 18.14 MB & 15.0 & 254.0 & n/a \\
  W2 & DeepProve 270M, 256+512 tok & n=2$^{\dagger}$19\% & 1465 & 4.53 & 18.13 MB & 22.9 & 254.1 & n/a \\
\bottomrule
\end{tabular}
\par\vspace{2pt}
\raggedright\scriptsize Baseline denominators are the median of the unmodified PyTorch step repetitions matched on workload, model, input length or batch, and machine. CPU rows ran on the host CPU only; ``--'' marks quantities the prover does not report.
\end{table*}

Fig.~\ref{fig:breakdown} shows the cost shares for the reported configurations, and Table~\ref{tab:overheads} summarizes the overheads of a representative subset. DeepProve's timers cannot separate its requantization and activation lookups from the sumcheck inside them, so 45--63\% of its GPU time is \emph{mixed}. Its cells run the CUDA backend, whereas its author-reported rates are CPU-only, so they serve as a breakdown rather than a rate comparison; one CPU cell at the author-reported setting provides the like-for-like ratio in Section~\ref{sec:published}.

\textbf{Q1 --- commitment and opening dominate only sometimes.} On VeriLoRA, opening is the largest category at short input lengths (46.7\% at input length 64 for 7B, 42.7\% for 13B) but falls to 26.1\% (24.5\%) at 2048, while lookup stays at 27--39\%. On zkLLM, opening leads at input length 1024 (35.9\%) but is overtaken by witness generation at 2048 (25.5\% versus 44.0\%). On DeepProve's GPU cells, commitment takes 18--25\% and opening only 5--9\%. On the Expander block, commitment and opening together are 1.6--5.5\% under all three schemes and the chained sumchecks take 80--85\%; on ZKTorch, commitment is the largest category (56.3\%); on VeriDP it is near zero. What dominates is a function of proof system, workload and scale.

\textbf{Q2 --- the witness share varies with workload and scale.} The witness share runs from effectively zero (Expander block, VeriDP) through 3--22\% (LoRA, DeepProve on GPU) and 14\% (ZKTorch) to 32--44\% (zkLLM). Where input length varies, it mostly grows with it: by 12 points from 1024 to 2048 on zkLLM, and from 3.6\% to 16.0\% (7B) and 4.4\% to 18.8\% (13B) between 64 and 2048 on VeriLoRA, consistent with the quadratic attention witness (B1) outgrowing the weight-related work. DeepProve is the exception at W1, where the share stays at 3--6\% between 64 and 1024 while its mixed category dominates; at W2 it grows with the number of generated tokens, from 6\% to 18\% between 32 and 512 at a prompt of 64.

\textbf{Q3 --- fixed-weight openings remain stable in the LoRA measurements.} The clearest evidence is VeriLoRA: across input lengths 64--2048, opening the weights takes approximately 11.5\,s (7B) or 12.4\,s (13B), varying by less than 1\%, while opening the intermediate tensors grows from 6.2 to 54.6\,s and from 6.7 to 67.7\,s. zkLLM shows a related shift in cost shares: doubling the input length raises proving time by 1.42$\times$ (33.6 to 47.8\,s) yet lowers the opening share by 10 percentage points. The commitment \emph{scheme}, in contrast, shows up mainly in proof size and verification: on the same block, moving from elliptic-curve commitments to the hash-based one grows the proof about 7$\times$ (0.90--0.96 to 6.56\,MB) and cuts verification by about 60\% (5.8--6.1 to 2.4\,s), while prover time falls by a third (13.5--14.2 to 9.3\,s). The two options also run on different fields (BN254 versus M31$\times$16), so this compares EC-commitment-on-BN254 with hash-commitment-on-M31, not the schemes alone.

\textbf{Q4 --- memory binds as a capacity wall before it binds as a time share.} Where it is measured, data movement is at most 8\% of proof time. Capacity is different. The 13B LoRA step at input length 2048 needs 45.0\,GB by the prover's own account (56.2\,GB sampled externally), so this configuration exceeds the 40\,GB card's capacity but completes in 327\,s on the 80\,GB device. The 7B step at the same length peaks at 39.9\,GB on the 80\,GB device, close to the smaller card's nominal capacity. DeepProve aborted during model commitment on the 40\,GB card, driven by vocabulary-embedding size rather than input length, but completes on the 80\,GB card with GPU peaks of 13--66\,GB; its 7B model failed on every configuration (buffer-allocation failures, or a shape-inference defect that also fails on CPU). Host memory is a wall too: DeepProve's host peak grows from 60\,GB at input length 64 to 260\,GB at 1024, and ZKTorch's largest verifiable model needs about 48\,GB on a CPU-only run.

\textbf{One-off and verification costs are not negligible.} Verification is 14--31\% of proving time on VeriLoRA and 26--43\% on the Expander block, a few seconds on DeepProve, and near zero on the interactive zkLLM. One-off setup is large where it exists: 295--460\,s on DeepProve, longer than the proof of a short context (270\,s at input length 64), and about a fifth of wall-clock on ZKTorch---the amortization opportunity S1 names.

\textbf{Answers.} No single category dominates across systems. Three generalizations hold in our data: each cell's cost concentrates in two or three categories; witness work grows with input length while weight-related opening does not (S1, S4); and memory binds first as capacity, on device and host, rather than as a share of time (B4). These trends guide the hardware assessment in Section~\ref{sec:hw}.

\subsection{How Much Can Engineering Remove?}
\label{sec:engineering}

\begin{figure}[t]
\centering
\includegraphics[width=\columnwidth]{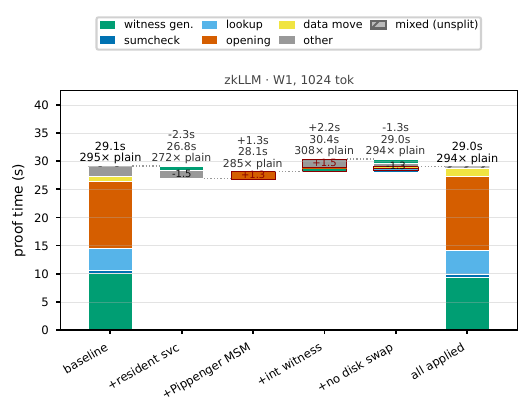}
\caption{Engineering-optimization waterfall for zkLLM on W1 (7B, one layer, input length 1024, A100-80GB, exclusive server). Each middle bar shows the per-category change against the previous bar; red-edged segments are regressions. Labels give the median proof time (one-off setup and verification excluded) and the slowdown over the same step in unmodified PyTorch on the same machine ($\times$plain). Medians of items 4 and 5 pool repetitions from the shared and the exclusive server (item 4 alone on the exclusive server: 30.8\,s).}
\label{fig:waterfall}
\end{figure}

We evaluate how implementation choices change the cost profile before considering hardware acceleration. Starting from a re-run baseline, we apply four changes to zkLLM in a fixed order: a resident prover service (no per-stage process start-up or CUDA initialization), Pippenger MSM on the opening path, integer-resident witness generation, and keeping intermediates on device instead of swapping to disk. Every version must pass a correctness gate before it is timed: an unmodified run verifies, and truncated inputs, corrupted activations or truncated weight files are rejected. (In this refactored code base the weight \emph{values} are not bound to the proved claim, so flipping weights alone still verifies---a limitation of the code, not of the gate.)

Fig.~\ref{fig:waterfall} shows a nearly unchanged aggregate proving time, from 29.1\,s to 29.0\,s, across this optimization sequence. The 29.1\,s baseline is a re-run on the exclusive 80\,GB server and therefore differs from the 33.6\,s measured on the shared 40\,GB server in Table~\ref{tab:overheads}. Individual steps have opposing effects, and the last two plotted medians pool repetitions from both servers, as stated in the caption. The resident service saves 2.3\,s of proving time and 18\% of wall-clock (34.8 to 28.5\,s); removing the disk swap saves 1.3\,s. Pippenger MSM adds 1.3\,s on both servers, even after tuning: a first version with 8192 bucket groups per window starved the GPU, and 1024 groups, the setting used in the timed runs, recovered most of that loss. Against the re-run baseline, however, the three large openings are unchanged (2.30 to 2.26--2.29\,s) while the three small ones become 33\% slower (1.27 to 1.69\,s), so the fixed cost of the bucket method outweighs its gain on small MSMs. Per-window parallelism, not the algorithm label, decides whether an MSM kernel helps. Integer-resident witness generation adds 2.2\,s on the exclusive server, whereas on the shared server the same steps saved 4.8\,s, consistent with, but not sufficient to isolate, an effect of disk contention; built with a newer CUDA toolchain it even failed its correctness check. The observed effects depend on the platform, configuration and toolchain. This sequence identifies changes that need further evaluation; it does not establish a bound on software optimization or on the gains attainable from hardware.

\subsection{Comparison with Published Systems}
\label{sec:published}

Table~\ref{tab:tasks} lists the author-reported costs of the published systems, normalized per token or per step. Within its caveats, per-token proving costs span about 0.02--0.9\,s across very different designs, verification stays between milliseconds and seconds, and the two systems that report a plaintext baseline put the slowdown at $10^{1}$--$10^{3}\times$ (DeepProve 24$\times$ to $10^{3}\times$~\cite{deepprove}, VeriLoRA about $10^{3}\times$~\cite{verilora})---consistent with the 249--341$\times$ we measure for one proved layer on W1 and the 92--143$\times$ on the small W4 networks.

Two reference comparisons illustrate the effects of configuration and implementation differences. Our zkLLM run---the refactored public release, interactive and without Fiat--Shamir, on synthetic fixed-point weights and a PCIe rather than SXM4 card---takes 47.8\,s per layer at input length 2048, or about 25.5 minutes for 32 layers, roughly 1.9$\times$ the author-reported 803\,s for a 13B model. For DeepProve one like-for-like ratio is available: the author-reported 64 tokens per minute use a 24-core CPU server with the HyperKZG commitment at input length 512, and the same setting on our CPU host, pinned to 24 threads, proves the same 0.27-billion-parameter model at 57.1 tokens per minute (median of three repetitions, 0.5\% spread), 0.89$\times$ the reported rate. Because model sizes, versions and hardware differ, the zkLLM extrapolation provides context rather than a matched replication. A lookup-cost study across two other provers~\cite{lookupsep} and vendor-reported figures~\cite{attestable,zkpytorch,joltatlas} complement this picture; we deliberately do not rank systems, since the purpose of this section is to show \emph{where} the cost goes, not which prover is fastest.

\section{State of Hardware Acceleration}
\label{sec:hw}

This section reviews acceleration on GPUs, FPGAs, ASICs and AI accelerators, and then assesses the body of work through the framework. Table~\ref{tab:accel} summarizes twenty representative designs. Their speed-ups are measured against different baselines and are not comparable across rows; what matters for our purpose is \emph{what} each design accelerates.

\begin{table*}[t]
\centering
\caption{Representative hardware acceleration of zero-knowledge provers. Speed-ups are as reported by the authors against the baselines named in parentheses and are not comparable across rows. The last column marks whether a neural-network model or circuit is part of the evaluation.}
\label{tab:accel}
\footnotesize
\setlength{\tabcolsep}{4pt}
\renewcommand{\arraystretch}{1.12}
\begin{tabular}{@{}l l >{\raggedright\arraybackslash}p{3.1cm} >{\raggedright\arraybackslash}p{3.0cm} >{\raggedright\arraybackslash}p{4.4cm} c@{}}
\toprule
\textbf{Work} & \textbf{Platform} & \textbf{Target proof system} & \textbf{Accelerated kernels} & \textbf{Reported speed-up (baseline)} & \textbf{AI} \\
\midrule
GZKP~\cite{gzkp}           & GPU & pairing-based SNARK & MSM, NTT & up to 48.1$\times$ (MINA, V100); up to 17.6$\times$ (bellperson, Zcash, four V100) & -- \\
cuZK~\cite{cuzk}           & GPU & pairing-based SNARK & MSM & MSM up to 2.94$\times$, overall up to 4.86$\times$ (prior GPU systems) & -- \\
BatchZK~\cite{batchzk}     & GPU & sumcheck + linear-code PCS & sumcheck, Merkle tree, encoder & $>$259.5$\times$ throughput (GPU systems); 9.52 proofs/s for verifiable ML & \checkmark \\
ZKPoG~\cite{zkpog}         & GPU & Plonkish, incl.\ witness generation & end-to-end & 22.8$\times$ (CPU), 12.7$\times$ (GPU) & -- \\
Chimera~\cite{chimera}     & GPU+CPU & Spartan & sumcheck & 7.96$\times$ sumcheck region; 1.27--1.64$\times$ end-to-end & -- \\
TensorZKP~\cite{tensorzkp} & GPU tensor cores & HyperPlonk + Brakedown & sumcheck, encoder, field arithmetic & 955$\times$ (CPU), 36.2$\times$ (SIMT GPU) & -- \\
\midrule
if-ZKP~\cite{ifzkp}        & FPGA (Intel) & pairing-based SNARK & MSM & 110--150$\times$ (software library) & -- \\
OPTIMSM~\cite{optimsm}     & FPGA (U55C) & any EC-based & MSM & up to 12.77$\times$ (prior FPGA) & -- \\
Hirner et al.~\cite{hashpcs_bw} & FPGA (U280, HBM) & Brakedown / Orion PCS & encoder, Merkle tree & commitment up to 264$\times$, proving up to 65$\times$ (Orion software, CPU); 3.85$\times$ end-to-end with Virgo & -- \\
\midrule
LegoZK~\cite{legozk}       & reconfig.\ ASIC (simulated) & multiple & MSM, NTT (shared field-arithmetic units) & avg.\ 31.96$\times$ (PipeZK), 11.30$\times$ (GZKP, V100) & \checkmark \\
ZK-Flex~\cite{zkflex}      & reconfig.\ accel. & multiple (POLY + EC) & NTT, MSM, multi-precision mult. & 5--11$\times$ (state of the art) & \checkmark \\
\midrule
PipeZK~\cite{pipezk}       & ASIC & pairing-based SNARK & NTT, MSM & 10$\times$ on standard benchmarks, 5$\times$ on Zcash (CPU); modules up to 197.5$\times$ (NTT), 77.7$\times$ (MSM) & -- \\
UniZK~\cite{unizk}         & ASIC & hash-based (Plonky2) & NTT, hashing, polynomial ops (unified) & avg.\ 97$\times$ (CPU), 46$\times$ (GPU) & -- \\
NoCap~\cite{nocap}         & ASIC & hash-based (Orion-type) & full prover & gmean 586$\times$ (32-core CPU); 41$\times$ (PipeZK, iso-resource) & -- \\
SZKP~\cite{szkp}           & ASIC & pairing-based SNARK & NTT, MSM (full proof) & $>$400$\times$ (CPU), 12$\times$ (GPU), 3$\times$ (ASIC) & -- \\
zkSpeed~\cite{zkspeed}     & ASIC & HyperPlonk & sumcheck, MSM & 801$\times$ geomean (CPU) & -- \\
zkPHIRE~\cite{zkphire}     & ASIC & HyperPlonk, high-degree gates & sumcheck, MSM & 1486$\times$ (CPU), 11.87$\times$ iso-area (SOTA) & -- \\
MTU~\cite{mtu}             & ASIC unit & sumcheck / Merkle-based & sumcheck, Merkle tree & up to 1478$\times$ (bandwidth-matched CPU) & -- \\
PipeSC~\cite{pipesc}       & ASIC & multilinear SNARKs & sumcheck & up to 2756$\times$ (Arkworks CPU), 5.02$\times$ (BatchZK GPU) & -- \\
\midrule
MORPH~\cite{aiasic}        & TPU v6e & EC-based & MSM, NTT & up to 10$\times$ NTT throughput, MSM on par (vs.\ gZKP, HPCA'23) & -- \\
\bottomrule
\end{tabular}

\par\vspace{2pt}
\raggedright\scriptsize
Baselines are given in parentheses. AI: \checkmark{} = a neural-network model or circuit is part of the evaluation (a model for BatchZK; circuits for ZK-Flex and for LegoZK, which reports LeNet, AlexNet and VGG16 circuits); -- = none reported.
\end{table*}

\subsection{GPUs: The Workhorse}
\label{sec:hw-gpu}

GPUs run almost every prover that reaches the scale of large language models today~\cite{zkllm,deepprove,verilora}, and AI providers already own them. Early work such as GZKP~\cite{gzkp} and cuZK~\cite{cuzk} parallelized MSM and NTT for pairing-based SNARKs and reduced host--device traffic, and open libraries now package these kernels~\cite{icicle,sppark}. A systematic characterization of GPU provers exposes the limit of this kernel-by-kernel approach: once MSM is optimized, NTT dominates proving latency~\cite{zkprophet}; its attribution of NTT to the 32-bit integer pipeline, however, is not borne out by our measurements (Section~\ref{sec:kernels}).

More recent systems widen the scope. BatchZK~\cite{batchzk} pipelines sumcheck, Merkle-tree and encoder modules so that many proofs are generated concurrently, overlapping transfers with computation. ZKPoG~\cite{zkpog} brings witness generation onto the GPU with the rest of a Plonkish prover and manages memory so that large circuits fit on consumer cards. Chimera~\cite{chimera} keeps the sumcheck state resident on the GPU across rounds, but its end-to-end gain stays far below its sumcheck speed-up because the polynomial commitment becomes the next bottleneck. TensorZKP~\cite{tensorzkp} uses the tensor cores that power AI, building finite-field arithmetic from low-precision matrix multiplication and recasting sumcheck and code encoding as matrix products.

GPUs are the default platform for verifiable AI. Their limits are the ones identified as B4: memory capacity, which decides whether a witness fits on one device, and the cost of moving data to the device, which falls hardest on streaming kernels such as encoding, sumcheck and field arithmetic (Section~\ref{sec:kernels}). Tensor cores open a path to exploit the numeric-format structure S3, but only for provers designed around them.

\subsection{FPGAs: Streaming and Efficiency}
\label{sec:hw-fpga}

FPGA work has concentrated on MSM: if-ZKP~\cite{ifzkp} scales Pippenger's algorithm across Intel FPGAs, and OPTIMSM~\cite{optimsm} decouples the number of buckets from the window size to save on-chip memory. Reconfigurable architectures aim at the whole prover: LegoZK~\cite{legozk} builds on the observation that every proving stage reduces to the same basic finite-field operations, and ZK-Flex~\cite{zkflex} combines a multi-precision multiplier core with a flexible on-chip network to keep utilization high as the workload shifts between polynomial and elliptic-curve stages.

The most instructive FPGA result for AI workloads concerns hash-based commitments. Hirner et al.~\cite{hashpcs_bw} show that for the commitment of Section~\ref{sec:why} it is random off-chip memory accesses, not arithmetic, that limit the encoder; by restructuring the code graph to remove random writes, generating it on the fly, and pipelining Merkle-tree hashing on an HBM-equipped FPGA, they accelerate the full commitment proof by about two orders of magnitude. This example illustrates how FPGAs offer energy efficiency and deep streaming pipelines. They are a poor fit for kernels whose speed is set by memory bandwidth, unless the dataflow is redesigned as in the example above, but they are well suited to hashing, to bit-level arithmetic over binary fields~\cite{binius}, and to processing data as it streams through the system.

\subsection{ASICs and AI Accelerators}
\label{sec:hw-asic}

ASIC designs specialize the datapath and memory hierarchy for proving workloads. PipeZK~\cite{pipezk} pipelined NTT and MSM for pairing-based SNARKs, and SZKP~\cite{szkp} targets full-proof acceleration. NoCap~\cite{nocap} and UniZK~\cite{unizk} target hash-based provers, the latter mapping their diverse kernels onto one array of vector-systolic units. A newer generation targets multilinear proof systems in which sumcheck dominates: zkSpeed~\cite{zkspeed} accelerates HyperPlonk end to end, zkPHIRE~\cite{zkphire} adds a programmable sumcheck unit for high-degree gates, MTU~\cite{mtu} exploits the binary-tree pattern shared by sumcheck and Merkle trees, and PipeSC~\cite{pipesc} pipelines sumcheck alone. Under equal memory budgets there is no universal winner between NTT- and sumcheck-based primitives~\cite{nttsumcheck}.

AI accelerators are an emerging alternative that providers already deploy at scale. MORPH~\cite{aiasic} converts high-precision modular arithmetic into dense low-precision matrix multiplications on TPU matrix units, exceeding a GPU baseline on NTT and matching it on MSM. Such reuse is attractive because proof systems evolve rapidly: within a few years the field has moved from univariate to multilinear constructions, from elliptic-curve to hash-based commitments, and from 256-bit to small fields~\cite{hyperplonk,basefold,binius}. A chip designed for one generation risks obsolescence before it ships, which favors programmable designs~\cite{zkphire,legozk,zkflex} and reuse of existing AI silicon~\cite{tensorzkp,aiasic}.

\subsection{A Workload-Aware Assessment}
\label{sec:hw-assess}

Viewed through the framework, the surveyed designs share four gaps with respect to AI workloads.

\textbf{Kernel focus lags behind AI proof systems.} Nine of the twenty designs in Table~\ref{tab:accel} take MSM or NTT as their main accelerated kernel, the kernels of pairing-based and univariate systems. Large-model provers, by contrast, are built on sumcheck and lookups (Section~\ref{sec:framework}). Sumcheck accelerators have appeared only recently~\cite{batchzk,zkspeed,zkphire,mtu,pipesc}, and none of the surveyed designs is built specifically around lookup arguments for non-linear layers.

\textbf{AI workloads are almost absent from evaluations.} Only three designs report an evaluation on a machine learning model or circuit: BatchZK~\cite{batchzk}, ZK-Flex~\cite{zkflex} and LegoZK~\cite{legozk}, all on convolutional networks. LegoZK's results for LeNet, AlexNet and VGG16 come from simulation of a 28\,nm design rather than from fabricated hardware. The others report generic benchmarks, with problem sizes of up to $2^{24}$ MSM points or $2^{30}$ constraints~\cite{optimsm,zkphire}. A forward pass of a 7-billion-parameter model over 2048 tokens already involves on the order of $10^{13}$ multiply--accumulate operations, so the memory behavior of AI-scale witnesses is largely untested.

\textbf{AI-specific structure is unexploited.} None of the surveyed accelerators is designed to reuse commitments to fixed weights across queries or to merge their openings (S1); batching of independent proofs~\cite{batchzk} is the closest instance of exploiting repetition (S2). Only one of them accelerates witness generation~\cite{zkpog}, although field-based witness generation accounts for up to 44\% of proving time in our zkLLM measurements (B1).

\textbf{End-to-end gains trail kernel gains.} When one kernel is accelerated, the bottleneck moves to the next (Section~\ref{sec:hw-gpu}), and our own waterfall shows that implementation-level changes to a prover can cancel each other out (Section~\ref{sec:engineering}). The end-to-end benefit of a kernel speed-up depends on its original cost share and on any new transfer or synchronization costs. This motivates joint design of witness generation, commitment, opening and data movement.

These gaps are the hardware counterpart of the gap between kernel speed-ups and end-to-end gains noted in the introduction. Section~\ref{sec:measure} quantifies them for concrete workloads, and Section~\ref{sec:open} turns them into open problems.

\section{Open Problems and Roadmap}
\label{sec:open}

The gaps identified in Sections~\ref{sec:measure} and~\ref{sec:hw} point to seven open problems. For each we note which bottleneck of the framework it addresses.

\textbf{1) Proof-friendly numeric formats (B1, S3).} Today a model is trained and served in floating point, then quantized and recomputed over a finite field for the proof. If the model itself ran in an integer or fixed-point format that the proof system can absorb, the witness could be produced as a by-product of ordinary execution on integer tensor units, removing most of the first bottleneck. Encouraging signs exist: integer-only quantization at 12 bits preserves the output of small language models closely~\cite{deepprove}, an industrial prover works directly on 8-bit integer matrix multiplication~\cite{attestable}, and tensor cores and TPU matrix units can already carry finite-field arithmetic~\cite{tensorzkp,aiasic}. Open questions are whether such formats preserve accuracy for training as well as inference, and how to keep execution bit-exact across hardware so that the proved computation is the one actually run~\cite{optimistic,bitexact}.

\textbf{2) Commit first, prove later (S1, B2).} Proving every inference or training step is expensive; committing to the computation is much cheaper. A provider could commit to intermediate states while it computes and produce proofs only on demand or for randomly sampled steps. A system that commits to block-boundary tensors during fine-tuning and later re-executes sampled blocks in trusted hardware reports 11--36\% training overhead across its evaluated models and stores 179--304\,GB per 2048 samples for its 14-billion-parameter model~\cite{aftune}, and a research agenda for verifying frontier training proposes real-time Merkle commitments as a core component~\cite{frontierzk}. Replacing the trusted re-execution with zero-knowledge proofs raises three questions: how cheap the always-on commitment can be made, what to store versus recompute, and how sampling rates translate into detection guarantees.

\textbf{3) Commitment on the data path (B4).} In distributed training and sharded inference, tensors already flow over the network between devices. Hashing or committing to them as they pass through SmartNICs or FPGAs would take commitment off the GPU's critical path and turn data movement from a cost into an opportunity. Hash-based commitments suit such streaming hardware well (Section~\ref{sec:hw-fpga}). A recent blueprint for verifying frontier training already relies on observing inter-node traffic through a physical tap or an attested SmartNIC and lists line-rate network taps among its open problems~\cite{frontierzk}; we are not aware of a published design that commits to tensors on the data path.

\textbf{4) Heterogeneous scheduling (B3, B4).} Table~\ref{tab:kernel-hw} summarizes the different resource demands of the kernels. Our GPU measurements distinguish compute-bound MSM, issue-bound low-degree sumcheck, synchronization- and issue-bound lookup, and bandwidth-bound encoding. Splitting a prover across devices---as already done between CPU and GPU for small sumcheck rounds~\cite{chimera}---could raise utilization, but every split adds transfers of large tables. A cost model that accounts for both kernel fit and data movement, and schedulers that use it, are missing.

\textbf{5) Co-selecting the commitment scheme and the hardware (B2, B3).} Elliptic-curve commitments concentrate cost in MSM, which suits GPUs and ASICs; hash-based commitments concentrate it in encoding and hashing over small fields, which suits FPGAs and bit-oriented logic~\cite{hashpcs_bw,binius}. Because the interactive part of a proof and its commitment are largely independent (Section~\ref{sec:layers}), the commitment scheme can be chosen to match the available hardware. The block-level comparison in Section~\ref{sec:breakdown} provides a starting point. Extending it to full AI workloads while controlling field and security parameters would help separate scheme effects from implementation effects.

\textbf{6) Security and performance are coupled.} Faster proofs must not weaken what is proved. A provider can keep a model's architecture and parameter count while choosing weights that collapse the computation into a much smaller model, and the proof still verifies~\cite{hollowllm}: a proof of correct inference is not a proof that a large model was executed. Every acceleration that changes the computation---lower-precision formats, sampling, merged openings---should come with an argument that the semantic guarantee the user cares about is preserved. Hardware cannot fix this alone, but hardware--protocol co-design must keep it in view.

\textbf{7) Benchmarks for AI-scale verifiable computation.} Accelerators are evaluated on generic circuits, and AI provers on their own models and hardware (Section~\ref{sec:hw-assess}). A shared benchmark suite---fixed models, input lengths and training steps, a common cost taxonomy, and reference implementations---would let hardware designers optimize for the workloads that matter. The workloads and taxonomy of Section~\ref{sec:measure} are offered as a starting point.

\begin{figure}[t]
\centering
\includegraphics[width=\columnwidth]{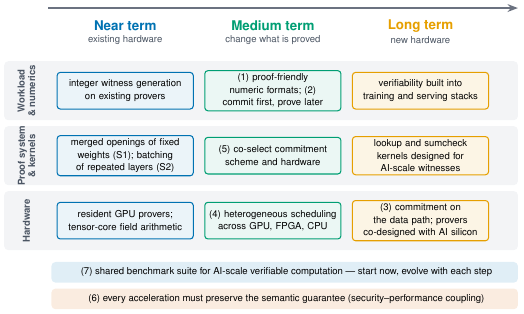}
\caption{Roadmap toward practical verifiable AI. Numbers refer to the open problems in Section~\ref{sec:open}.}
\label{fig:roadmap}
\end{figure}

\textbf{Roadmap.} Fig.~\ref{fig:roadmap} presents these problems by time horizon. In the near term, existing GPUs and shared benchmarks support evaluations of software optimizations, merged openings of fixed weights, batching of repeated layers and field arithmetic on tensor cores. The mixed effects in Section~\ref{sec:engineering} make end-to-end validation on the target platform essential. In the medium term, proof-friendly numeric formats, commit-first-prove-later training and heterogeneous scheduling change what has to be proved and where. In the long term, commitment on the data path and accelerators co-designed with AI silicon could make verifiability a built-in property of AI infrastructure rather than an expensive add-on.

\section{Conclusion}
\label{sec:conclusion}

This article connects the cost of verifiable AI to hardware design through a four-layer framework spanning verifiable AI workloads, proof systems, cryptographic kernels and platforms. Cross-system measurements show that acceleration priorities change with the prover and input length: witness generation, commitment and opening, lookup and sumcheck each dominate in different configurations. The LoRA measurements identify nearly constant fixed-weight opening costs alongside growing work on intermediate tensors, while the 13-billion-parameter configuration exposes a capacity boundary between 40 and 80\,GB devices. Kernel profiling further shows that instruction issue and synchronization can limit kernels commonly described as bandwidth-bound. Together, these findings make workload-specific cost profiles and memory requirements central to accelerator selection.

The hardware review identifies opportunities to connect recent kernel advances to AI-scale evaluations, fixed-weight reuse and witness generation. Software optimization remains part of this agenda: the zkLLM case study shows opposing effects across a particular sequence of changes and motivates validation on the intended platform. Proof-friendly numeric formats, coordinated commitment and opening, heterogeneous execution and shared benchmarks provide a path toward integrating proof generation into AI infrastructure.

% \section*{Acknowledgment}
% AI-assisted tools, including OpenAI Codex~\cite{openai_codex} and Anthropic Claude Code~\cite{anthropic_claude_code}, were used to assist with drafting and revising prose in the Abstract and Sections~I--VII and with developing and checking scripts for parsing, aggregating, and formatting raw experimental outputs. The underlying measurements were obtained from the authors' experimental runs. All AI-assisted scripts, processed data, and derived tables were reviewed and validated against the raw logs by the authors, who take full responsibility for the reported content and results.

\bibliographystyle{IEEEtran}
\bibliography{refs}

\end{document}